\documentclass[aps,prd,showpacs,eqsecnum,twocolumn,superscriptaddress,nofootinbib]
{revtex4}
\usepackage{amsmath,amssymb,graphicx,color}

\begin{document}

\title{Exploring tidal effects of coalescing binary neutron stars 
in numerical relativity II: Longterm simulations}

\author{Kenta Hotokezaka}
\affiliation{Racah Institute of Physics, The Hebrew University of Jerusalem, Jerusalem, 91904, Israel} 

\author{Koutarou Kyutoku} 
\affiliation{Department of Physics,
University of Wisconsin-Milwaukee, P.O. Box 413, Milwaukee,
Wisconsin 53201, USA}

\author{Hirotada Okawa}
\affiliation{Yukawa Institute for Theoretical Physics, 
Kyoto University, Kyoto, 606-8502, Japan} 
\affiliation{Advanced Research Institute for Science \&
Engineering, Waseda University, 3-4-1 Okubo,
Shinjuku, Tokyo 169-8555, Japan}  

\author{Masaru Shibata}
\affiliation{Yukawa Institute for Theoretical Physics, 
Kyoto University, Kyoto, 606-8502, Japan} 

\date{\today}

\newcommand{\beq}{\begin{equation}}
\newcommand{\eeq}{\end{equation}}
\newcommand{\beqn}{\begin{eqnarray}}
\newcommand{\eeqn}{\end{eqnarray}}
\newcommand{\pa}{\partial}
\newcommand{\vp}{\varphi}
\newcommand{\varep}{\varepsilon}
\newcommand{\ep}{\epsilon}
\newcommand{\comp}{(M/R)_\infty}
%%%%%%%%%%%%%%%%
\begin{abstract}
%%%%%%%%%%%%%%%% 

We perform new longterm (15--16 orbits) simulations of coalescing
binary neutron stars in numerical relativity using an updated
Einstein's equation solver, employing low-eccentricity initial data,
and modeling the neutron stars by a piecewise polytropic equation of
state. A convergence study shows that our new results converge more
rapidly than the third order and using the determined convergence
order, we construct an extrapolated waveform for which the estimated
total phase error should be less than 1 radian. We then compare the
extrapolated waveforms with those calculated by the latest
effective-one-body (EOB) formalism in which the so-called tidal
deformability, higher post-Newtonian corrections, and gravitational
self-force effects are taken into account. We show that for a binary
of compact neutron stars with their radius 11.1\,km, the waveform by
the EOB formalism agrees quite well with the numerical waveform so
that the total phase error is smaller than 1~radian for the total
phase of $\sim 200$~radian up to the merger. By contrast, for a binary
of less compact neutron stars with their radius 13.6\,km, the EOB and
numerical waveforms disagree with each other in the last few wave
cycles, resulting in the total phase error of $\sim 3$~radian.

%%%%%%%%%%%%%%
\end{abstract}
%%%%%%%%%%%%%%

\pacs{04.25.D-, 04.30.-w, 04.40.Dg}

\maketitle

\section{Introduction}

The inspiral and merger of coalescing compact binaries are among the
most promising sources for ground-based kilometer-size
laser-interferometric gravitational-wave detectors~\cite{Detectors}.  A
statistical study based on the stellar-evolution synthesis (e.g.,
Refs.~\cite{O'S10,abadie10}) suggests that the detection rate for them
will be $\sim 1$--$100\,\rm{yr}^{-1}$ for advanced detectors, i.e.,
advanced LIGO~\cite{aligo}, advanced VIRGO~\cite{avirgo}, and
KAGRA~\cite{kuroda10}, which will sequentially start operation in the
coming years.

One of the important steps after the first detection of gravitational
waves from binary neutron stars (and also a black hole-neutron star
binary) will be to extract binary parameters such as mass, spin, and
radius of each object in the binary systems.  In particular, the mass
and radius (or a quantity related to it) of the neutron stars have
invaluable information for determining the equation of state (EOS) of
the neutron-star matter, which is still poorly known.  The mass of two
neutron stars will be determined with a high accuracy $\alt 1\%$, if
the gravitational-wave signals in the inspiral stage are detected with
the signal-to-noise ratio $\agt 10$ and the neutron-star spins are
supposed to be negligible~\cite{cutler94}. On the other hand,
determining the parameters related to the neutron-star radius is the
challenging issue although it has to be done for constraining the
neutron-star EOS~(e.g., Ref.~\cite{read13,bauswein14,takami14}).
Among other possible methods, extracting the {\em tidal deformability}
of the neutron stars from gravitational waves emitted
binary-neutron-star inspirals is one of the most promising
methods~\cite{lai94,flanagan08,BDF2012}. For employing this method, we
have to prepare a theoretical template of gravitational waves from
binary-neutron-star inspirals taking into account tidal-deformation
effects that influence the dynamics of the late inspiral orbits (e.g.,
Ref.~\cite{mora04}).  Hence, theoretically deriving a precise
gravitational waveform for binary-neutron-star inspirals including the
tidal effects is an urgent task.

A post-Newtonian (PN) gravitational waveform for the early stage of
binary-neutron-star inspirals (with the frequency $f \lesssim 400$~Hz)
was first derived by Flanagan and Hinderer including the leading-order
tidal effects~\cite{flanagan08}.  They showed that the tidal effect
for the evolution of the gravitational-wave phase could be described
only by the tidal deformability of neutron stars.  They also found
that the tidal deformability of neutron stars could be measured by the
advanced gravitational-wave detectors by using the gravitational-wave
signals for $f=10$ -- 400~Hz, if the tidal deformability of neutron
stars is sufficiently large or if we could observe an event with a
high signal-to-noise ratio (see also Ref.~\cite{hinderer10}).  If the
waveform is extended to the higher frequency range, the measurability
can be significantly improved. In the PN approach, however, the
uncertainty of the higher-order PN terms prevents us to construct the
accurate waveform at the higher
frequency~\cite{favata14,yagi14,wade14}.

To overcome the ambiguity in the higher PN terms, an effective one
body~(EOB) formalism with the tidal effects has been
explored~\cite{BDF2012,BD2014, bernuzzi14}. In this approach, the
non-tidal part is calibrated using the results of binary-black-hole
simulations.  Damour and his collaborators~\cite{damour12}
subsequently explored the measurability of the tidal deformability
with the advanced gravitational-wave detectors employing an EOB
formalism including tidal effects up to the second PN order.  They
concluded that the tidal deformability of neutron stars could be
measured by the advanced gravitational-wave detectors if the
signal-to-noise ratio of the gravitational-wave signal is higher than
$\sim 16$ for any EOS that satisfies the constraint of the maximum
gravitational mass $\agt 2M_{\odot}$~\cite{demorest10}.  The key
assumption of their study is that the EOB approach is valid up to the
onset of the merger of binary neutron stars.  However, in the stage
just before the merger, effects such as {\em nonlinear}
tidal-deformation effects, which are not taken into account in the
current EOB formalism, could come into play (see, e.g.,
Ref.~\cite{HKS2013}). 

%%In addition, higher-PN tidal corrections may
%%yield a pole-like nonlinear behavior of the tidal interactions near
%%the last unstable orbit~\cite{BDF2012}.

For precisely understanding the orbital motion and the waveform in the
late inspiral stage of binary neutron stars, a high-resolution
numerical-relativity (NR) simulation with appropriately physical
setting is obviously necessary.  Recently, long-term simulations for
binary-neutron-star inspirals were performed by three
groups~\cite{baiotti11,thierfelder11,bernuzzi11,bernuzzi12,HKS2013,radice14,bernuzzi14}
aiming at the derivation of accurate gravitational waveforms for the
late inspiral stage. They followed the late binary inspiral for $\alt
10$ orbits up to the onset of the merger.  However, in their numerical
simulations, an unphysical residual eccentricity is present in the
initial data. This seriously made their results less accurate, because
binary neutron stars in the late inspiral stage are believed to have a
quasi-circular orbit with negligible eccentricity.  In the present 
work, we simulate binary-neutron-star inspirals for a longer term with
more physical initial data in which the eccentricity is sufficiently
small (less than $10^{-3}$) \footnote{We note that R. Haas and his
  collaborators (SXS collaboration) have also derived the waveforms of
  small eccentricity in their longterm simulations, although their
  results have not been published yet.}.  In addition, we perform the
simulations employing a formalism in which the constraint violation
can be suppressed to a level much smaller than that in our previous
study~\cite{HKS2013}.  As a result, we can obtain an extrapolated
waveform in a much more accurate and reliable manner than in our
previous study.

The paper is organized as follows.  In Sec.~II, we summarize the
formulation and numerical schemes employed in our numerical-relativity
study, and also review the EOS employed.  In Sec.~III, we describe our
method for deriving an extrapolated gravitational waveform, showing the
resulting waveforms that are much more accurate than those derived in
our previous study~\cite{HKS2013}. We then compare our extrapolated
waveforms with those derived by the latest EOB approach and examine its
accuracy in Sec.~IV.  Section~V is devoted to a summary.  Throughout
this paper, we employ the geometrical units of $c=G=1$ where $c$ and $G$
are the speed of light and the gravitational constant, respectively.

\section{Formulation for numerical-relativity simulation}

In this section, we briefly describe the formulation and the numerical
schemes of our numerical-relativity simulation employed in this work. 

\subsection{Evolution and Initial Condition}

We follow the inspiral and early stage of the merger of binary neutron
stars using our numerical-relativity code, {\tt SACRA}, for which the
details are described in Ref.~\cite{yamamoto08}.  In this work, we
employ a moving puncture version of the
Baumgarte-Shapiro-Shibata-Nakamura formalism~\cite{BSSN}, {\em
  locally} incorporating a Z4c-type constraint propagation
prescription~\cite{Z4c} (see~\cite{KST2014} for our implementation)
for a solution of Einstein's equation. The constraint propagation from
the neutron-star's outer region plays a crucial role for reducing the
constraint violation and for improving the order of the convergence as
discovered in Ref.~\cite{Z4c}.  In our numerical simulation, a
fourth-order finite differencing scheme in space and time is used
implementing an adaptive mesh refinement (AMR) algorithm.  At
refinement boundaries, a second-order interpolation scheme is partly
used.  The advection terms are evaluated by fourth-order lop-sided
upwind-type finite differencing~\cite{brugmann08}.  A fourth-order
Runge-Kutta method is employed for the time evolution.  For the
hydrodynamics, a high-resolution central scheme based on a
Kurganov-Tadmor scheme~\cite{kurganov00} with a third-order piecewise
parabolic interpolation and with a steep min-mod limiter is employed.

In this work, we prepare nine refinement levels for the AMR
computational domain. Specifically, two sets of four finer domains
comoving with each neutron star cover the region of their vicinity.
The other five coarser domains cover both neutron stars by a wider
domain with their origins fixed at the center of the mass of the
binary system.  Each refinement domain consists of a uniform,
vertex-centered Cartesian grid with $(2N+1,2N+1,N+1)$ grid points for
$(x,y,z)$ (the equatorial plane symmetry at $z=0$ is imposed). The
half of the edge length of the largest domain (i.e., the distance from
the origin to outer boundaries along each axis) is denoted by $L$,
which is chosen to be larger than $\lambda_{0}$, where $\lambda_{0} =
\pi/\Omega_{0}$ is the initial wavelength of gravitational waves and
$\Omega_{0}$ is the initial orbital angular velocity. The grid spacing
for each domain is $\Delta x_{l}=L/(2^{l}N)$, where $l=0-8$.  In this
work, we choose $N=72$, 60, 48, and 40 for examining the convergence
properties of numerical results. With the highest grid resolution (for
$N=72$), the semimajor diameter of each neutron star is covered by
about 120 grid points.

\begin{table*}[t]
\centering
\caption{\label{tab1} Equations of state (EOS) employed, the radius and
  the tidal Love number of $l=(2,~3,~4)$ of
  spherical neutron stars of mass $1.35M_\odot$, the radius of light ring orbit,
  angular velocity of
  initial data, and the finest grid spacing in the four different
  resolution runs. $m_0$ denotes the total mass of the system.  In this
  study, it is $2.7M_\odot$.}
\begin{tabular}{ccccccccc}
\hline\hline
~~EOS~~ & ~~$R_{1.35}$\,(km)~~ & ~~$k_{2,1.35}$~~ & ~~$k_{3,1.35}$~~ 
& ~~$k_{4,1.35}$~~& ~~$r_{\rm LR}$~~ & $m_0\Omega_0$ & $\Delta x_{8}$\,(km) &   \\
\hline
APR4 & 11.1 & 0.0908 & 0.0234 & 0.00884 & 3.61 &~ 0.0156~  &~~ 0.140,~0.167,~0.209,~0.251 & \\
H4   & 13.6 & 0.115 & 0.0326 & 0.0133 & 4.21 &~ 0.0155~  &~~ 0.183,~0.220,~0.274,~0.329 & \\
\hline\hline
\end{tabular}
\end{table*}

We prepare binary neutron stars in quasi-circular orbits for the
initial condition of numerical simulations. These initial conditions
are numerically obtained by using a spectral-method library,
LORENE~\cite{lorene}. We follow 15--16 orbits in this study. To do so,
the orbital angular velocity of the initial configuration is chosen to
be $m_0\Omega_{0} \approx 0.0155$ ($f \approx 370$\,Hz for the total
mass $m_0=2.7M_{\odot}$, i.e., each mass of neutron stars is
$1.35M_\odot$).  The neutron stars are assumed to have an irrotational
velocity field, which is believed to be an astrophysically realistic
configuration~\cite{bildsten92,kochaneck92}.  The parameters for the
initial models are listed in Table~\ref{tab1}.

For the computation of an accurate gravitational waveform in numerical
simulations, we have to employ initial data of a quasi-circular orbit
of negligible eccentricity.  Namely, the eccentricity of the initial
binary orbit has to be reduced to be as small as possible. Such
initial data are constructed by an eccentricity-reduction procedure
described in~\cite{KST2014}.  For the initial data employed in this
work, the residual eccentricity is $\alt 10^{-3}$.  

\subsection{Equation of State}

Following previous works~\cite{HKS2013,KST2014}, we employ a
parameterized piecewise-polytropic equation of state proposed by
Read and her collaborators~\cite{read09}. This EOS is written in terms
of four segments of polytropes
\begin{align}
P = K_{i}\rho^{\Gamma_{i}} ~~~
\text{( for $\rho_{i}\leq \rho<\rho_{i+1}$, $0\leq i \leq 3$)},~~
\end{align}
where $\rho$ is the rest-mass density, $P$ is the pressure, $K_{i}$ is a
polytropic constant, and $\Gamma_{i}$ is an adiabatic index.  At each
boundary of the piecewise polytropes, $\rho=\rho_{i}$, the pressure is
required to be continuous, i.e.,
$K_{i}\rho_{i+1}^{\Gamma_{i}}=K_{i+1}\rho_{i+1}^{\Gamma_{i+1}}$. Following
Read and her collaborators, these parameters are determined in the
following manner~\cite{read09}: The crust EOS is fixed by setting
$\Gamma_{0}=1.3562395$ and $K_{0}=3.594\times 10^{13}$ in cgs units.  The
values of the boundary density is set as $\rho_{2}=10^{14.7}\,{\rm
g/cm^3}$ and $\rho_{3}=10^{15.0}\,{\rm g/cm^3}$.  With this preparation,
the following four parameters become free parameters that should be
given: $\{P_{1}, \Gamma_{1}, \Gamma_{2}, \Gamma_{3}\}$.  Here, $P_{1}$
is the pressure at $\rho=\rho_{2}$, and for a given value of this,
$K_{1}$ and $K_{i}$ are determined by
$K_{1}=P_{1}/\rho_{2}^{\Gamma_{1}}$ and
$K_{i+1}=K_{i}\rho_{i+1}^{\Gamma_{i}-\Gamma_{i+1}}$.
%%Therefore, the EOS is specified by choosing the four parameters
%%$\{P_{1}, \Gamma_{1}, \Gamma_{2}, \Gamma_{3}\}$.
In this work, we choose two sets of piecewise-polytropic EOS
mimicking APR4~\cite{APR} and H4 \cite{H4} EOS (see Table 1 of
Ref.~\cite{Hotokezaka2013} for the four parameters).

In numerical simulations, we employ a modified version of the
piecewise polytropic EOS to approximately take into account thermal
effects, which play a role in the merger phase.  In this EOS, we
decompose the pressure and specific internal energy into the cold and
thermal parts as
\begin{equation}
 P = P_{\rm cold}(\rho) + P_{\rm th}\,,~~~~\varepsilon = \varepsilon_{\rm
  cold}(\rho) + \varepsilon_{\rm th} .
\end{equation}
The cold parts of both variables are calculated using the original
piecewise polytropic EOS from $\rho$, and then the thermal part of the
specific internal energy is defined from $\varepsilon$ as
$\varepsilon_{\rm th} = \varepsilon - \varepsilon_{\rm
  cold}(\rho)$. Because $\varepsilon_{\rm th}$ vanishes in the absence
of shock heating, it is regarded as the finite-temperature part
determined by the shock heating in the present context.  For the
thermal pressure, a $\Gamma$-law ideal-gas EOS was adopted as
\begin{equation}
 P_{\rm th} = ( \Gamma_{\rm th} - 1 ) \rho \varepsilon_{\rm th}.
\end{equation}
Following our latest works~\cite{Hotokezaka2013,KST2014},
$\Gamma_{\rm th}$ is chosen to be 1.8. 

\subsection{Extraction of Gravitational waves}

Gravitational waves are extracted from the outgoing-component of
complex Weyl scalar $\Psi_{4}$~\cite{yamamoto08}. From this, 
gravitational waveforms are determined in spherical coordinates $(r ,
\theta , \phi )$ by
\begin{eqnarray}
h:=h_{+}(t)-ih_{\times}(t) = -\lim_{r\rightarrow \infty}
\int^{t}dt^{\prime}\int^{t^{\prime}}dt^{\prime \prime}
\Psi_{4}(t^{\prime \prime},r).\nonumber \\ \label{eq:hpsi4}
\end{eqnarray}
Here, we omit arguments $\theta$ and $\phi$. $\Psi_{4}$ can be expanded
in the form
\begin{eqnarray}
\Psi_{4}(t, r, \theta, \phi) = \sum_{lm}\Psi_{4}^{l,m}(t,r)_{-2}
Y_{lm}(\theta, \phi),
\end{eqnarray}
where $_{-2}Y_{lm}$ denotes the spin-weighted spherical harmonics of
weight $-2$ and $\Psi_{4}^{l,m}$ are expansion coefficients defined by
this equation.  In this work, we focus only on the $(l,|m|)=(2,2)$
mode because we pay attention only to the equal-mass binary, and
hence, this quadrupole mode is the dominant one.

%The gravitational-wave phase $\phi_{\rm{NR}}$ is defined by
%\begin{eqnarray}
%\Psi_{4}^{22}(t,r) = A_{22}(t,r)e^{i\phi_{\rm{NR}}(t,r)},
%\end{eqnarray}
%where $A_{22}$ denotes the amplitude and it is real.  

We evaluate $\Psi_{4}$ at a finite spherical-coordinate radius
$r/m_0=100$--240.  To compare the waveforms extracted at different
radii, we use the retarded time defined by
\begin{eqnarray}
t_{\rm{ret}} := t - r_{*}, \label{eq:tret1}
\end{eqnarray}
where $r_{*}$ is the so-called tortoise coordinate defined by
\begin{eqnarray}
r_{*} := r_{\rm{A}} + 2m_0\ln \left(\frac{r_{\rm{A}}}{2m_0}-1\right),
 \label{eq:tret2}
\end{eqnarray}
with $r_{\rm{A}}:=\sqrt{A/4\pi}$ and $A$ the proper area of the
extraction sphere.

\section{Recipe for constructing an extrapolated waveform}

In this section, we present our prescription for deriving an
extrapolated gravitational waveform from raw numerical data of
$\Psi_4$, and show that the resulting waveforms have a good accuracy 
that can be compared carefully with the EOB results.

\subsection{Extrapolation to infinite extraction radius}

%%%%%%%%%%%% MAY NEED FIGURE
\begin{figure*}[t]
\begin{center}
\includegraphics[width=84mm]{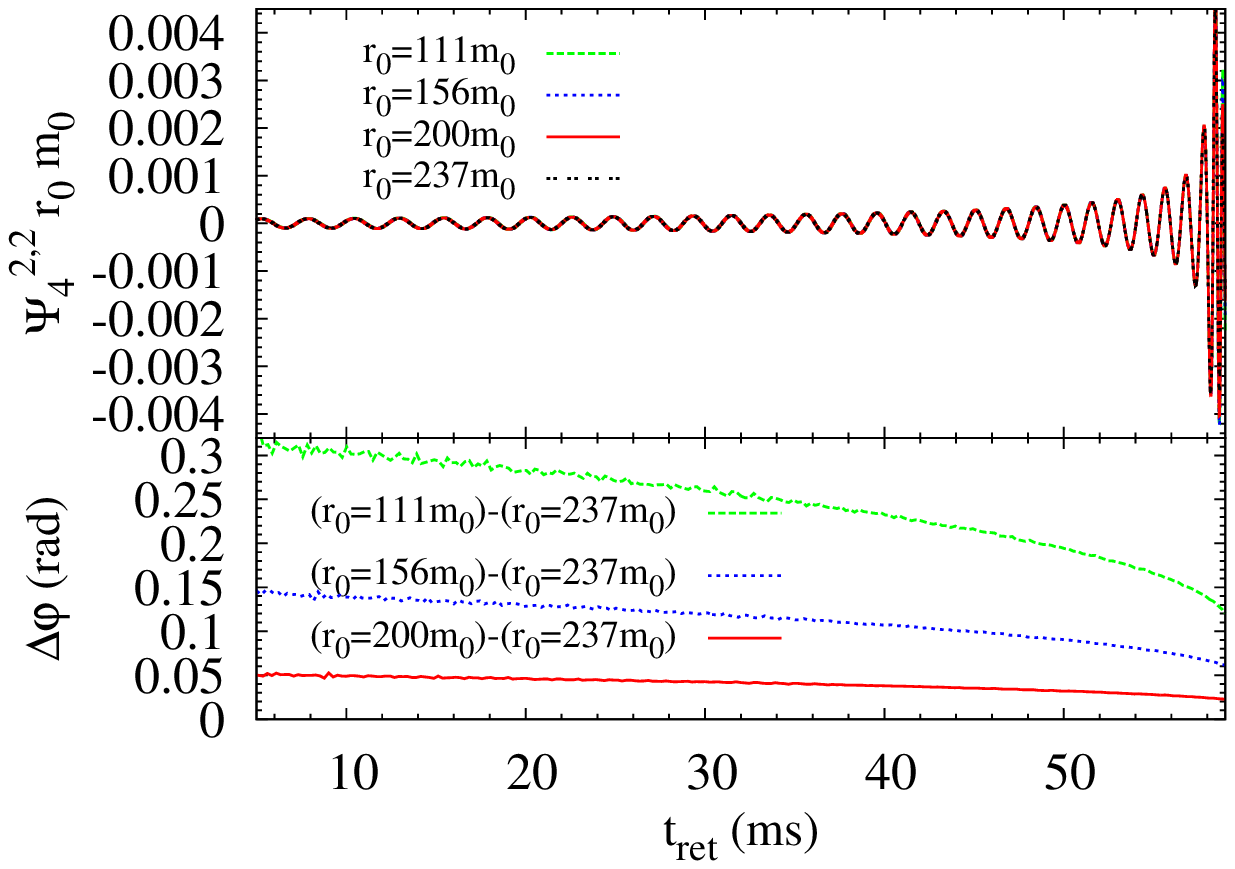}
~~~\includegraphics[width=84mm]{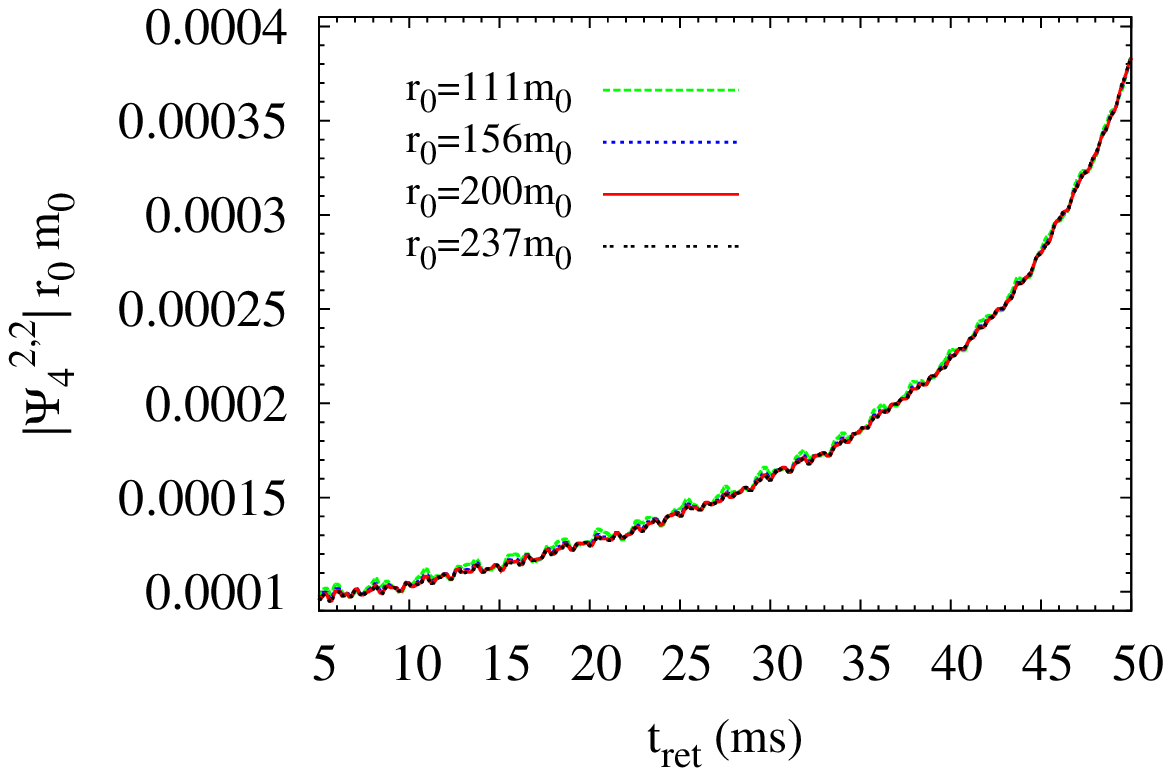}
%\vspace*{-4mm}
\caption{The waveform (real part; left) and amplitude (right) of
  $\Psi_4^{2,2,\infty}(r_0,t_{\rm ret})$ as functions of $t_{\rm ret}$
  for several values of $r_0$ for the run with H4 EOS and the best
  grid resolution ($N=72$). The lower plot of the left panel shows
  the phase differences of $\Psi_4^{2,2,\infty}(r_0)$ relative to
  $\Psi_4^{2,2,\infty}(r_0=237m_0)$. 
\label{fig1}}
\end{center}
\end{figure*}

As we mentioned in the previous section, we extract $\Psi_4$ at several
coordinate radii, $100$--$240m_0$, and then, this complex Weyl scalar is
decomposed into the spherical harmonics components, $\Psi_4^{l,m}$.
Since the waveform of $\Psi_4$ extracted at a finite radius, $r_0$, is
systematically different from that at null infinity, we first compute an
extrapolated waveform at $r_0 \rightarrow \infty$ using the Nakano's
method as~\cite{LNZC2010,Nakano15}
\beqn
\Psi_4^{l,m,\infty}(t_{\rm ret}, r_0)
&=&C(r_0)\left[\Psi_4^{l,m}(t_{\rm ret}, r_0) \right.\nonumber \\
&~&\left. -{(l-1)(l+2) \over 2r_{\rm A}}
\int^{t_{\rm ret}} \Psi_4^{l,m}(t', r_0)dt'
\right], \nonumber \\
\label{eq:nakano}
\eeqn
where $C(r_0)$ is a function of $r_0$.  Since our coordinates are
similar to isotropic coordinates of non-rotating black holes, we
choose $r_{\rm A}=r_0[1+m_0/(2r_0)]^2$.  $C(r_0)$ depends on the choice of
the tetrad components; for our choice, it is appropriate to choose
$C(r_0)=1 - 2m_0/r_{\rm A}$. In this setting, $t_{\rm ret}$ at $r=r_0$
is given by Eqs.~(\ref{eq:tret1}) and (\ref{eq:tret2}).
%%%

The left panel of Fig.~\ref{fig1} plots the real part of
$\Psi_4^{2,2,\infty}(t_{\rm ret}, r_0)$ for several choices of $r_0$.
The right panel shows the evolution of the absolute amplitude of
$\Psi_4^{2,2,\infty}(t_{\rm ret}, r_0)$.  These show that the
extrapolated waveforms depend very weakly on the extraction radius,
$r_0$ (see Ref.~\cite{Nakano15} for the reason).

We then have to calibrate how weakly the resulting extrapolated
waveforms, $\Psi_4^{2,2,\infty}(t_{\rm ret}, r_0)$, depend on $r_0$
and have to estimate the systematic error in this quantity.  We find
that the systematic error in phase decreases approximately in
proportional to $r_0^{-1}$ (cf.~the left lower panel of 
Fig.~\ref{fig1} that indeed shows this property). Figure~\ref{fig1}
implies that for $r_0 \agt 200m_0$, the systematic error in phase is
smaller than 0.3 radian. This value is smaller than the error in the
extrapolated waveform finally obtained (associated with the
uncertainty in the resolution extrapolation), and can be accepted in
the present numerical study. Note that this phase error is systematic
and could be subtracted, although we do not do so in this work.

By contrast, the systematic error in amplitude is appreciable, i.e.,
1--2 percents even for $r_0 \approx 200m_0$. For suppressing this error,
we might have to enlarge the computational domain for the wave
extraction.  However, this error size is smaller than another error
associated with the spurious short-term modulation in the numerical
gravitational-wave amplitude as reported in Ref.~\cite{KST2014}: The right
panel of Fig.~\ref{fig1} shows that a modulation in the amplitude is
present with its fluctuation amplitude of $\alt 3\%$ in particular in
the early stages of the numerical waveform.  Since this error was not
able to be cleaned up, we do not take a further extrapolation of
$|\Psi_4^{l,m,\infty}(t_{\rm ret}, r_0)|$ for $r_0 \rightarrow \infty$.
Thus, in this work, we employ $\Psi_4^{l,m,\infty}(t_{\rm ret}, r_0)$
computed from the data extracted at $r_0=200m_0$ [hereafter written as
$\Psi_4^{l,m,\infty}(t_{\rm ret})$] without further processing and
perform subsequent analyses keeping in mind that in the amplitude
extrapolated by Eq.~(\ref{eq:nakano}), there could exist a local error
in magnitude up to $\sim 3\%$ of the exact amplitude (note that in
average the error would be much smaller than 3\%).

\subsection{Extrapolation for zero-grid spacing limit}

\begin{figure*}[t]
\begin{center}
\includegraphics[width=84mm]{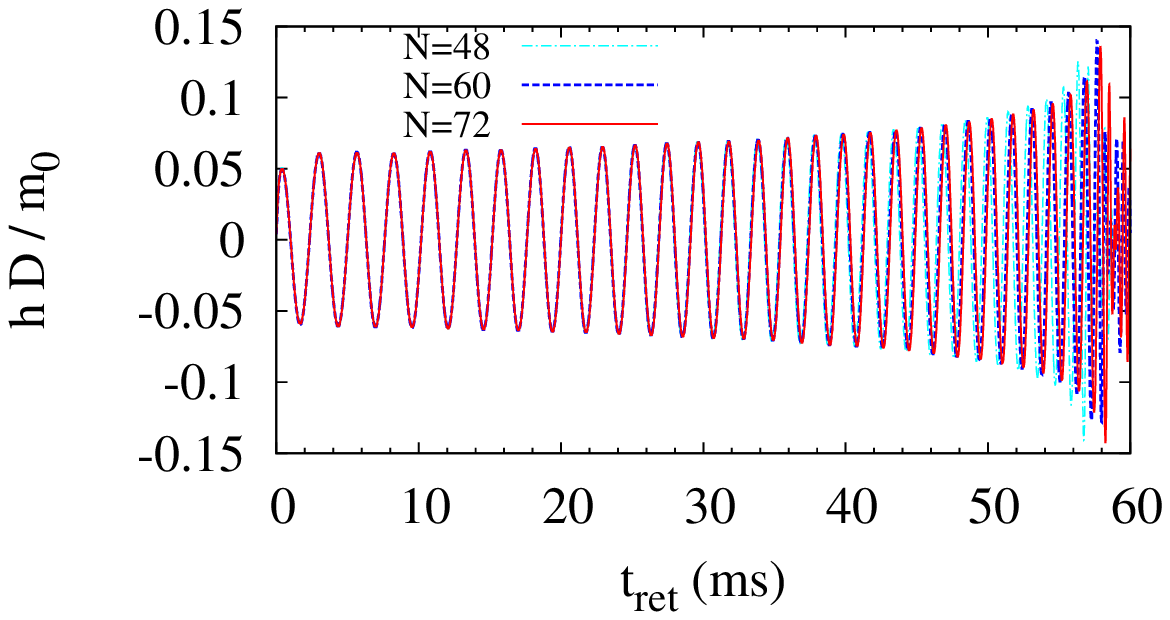}~~~
\includegraphics[width=84mm]{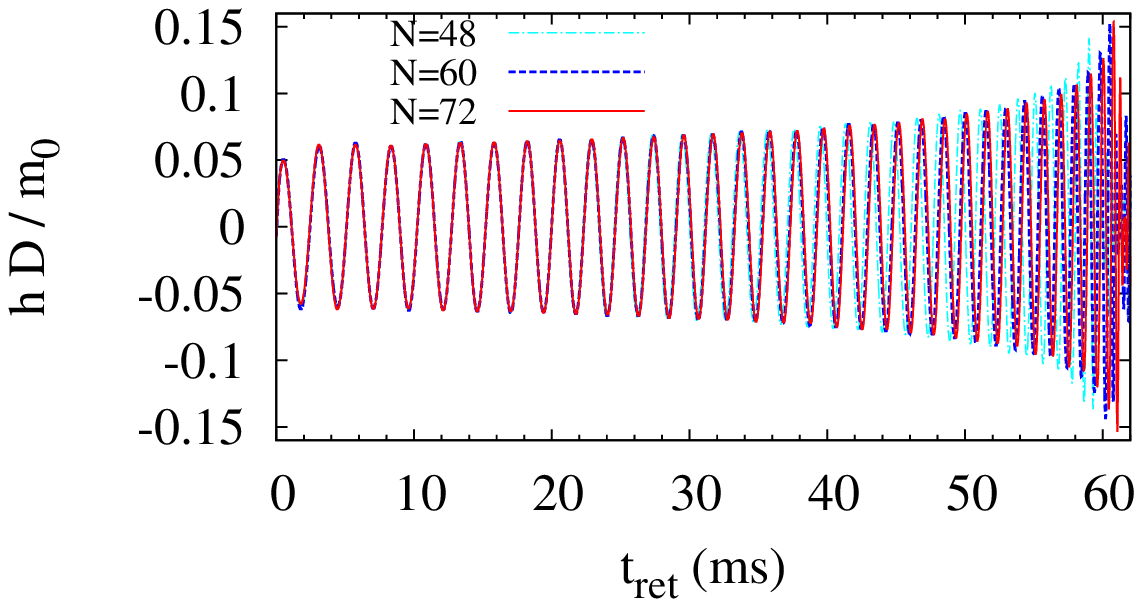}\\
\includegraphics[width=84mm]{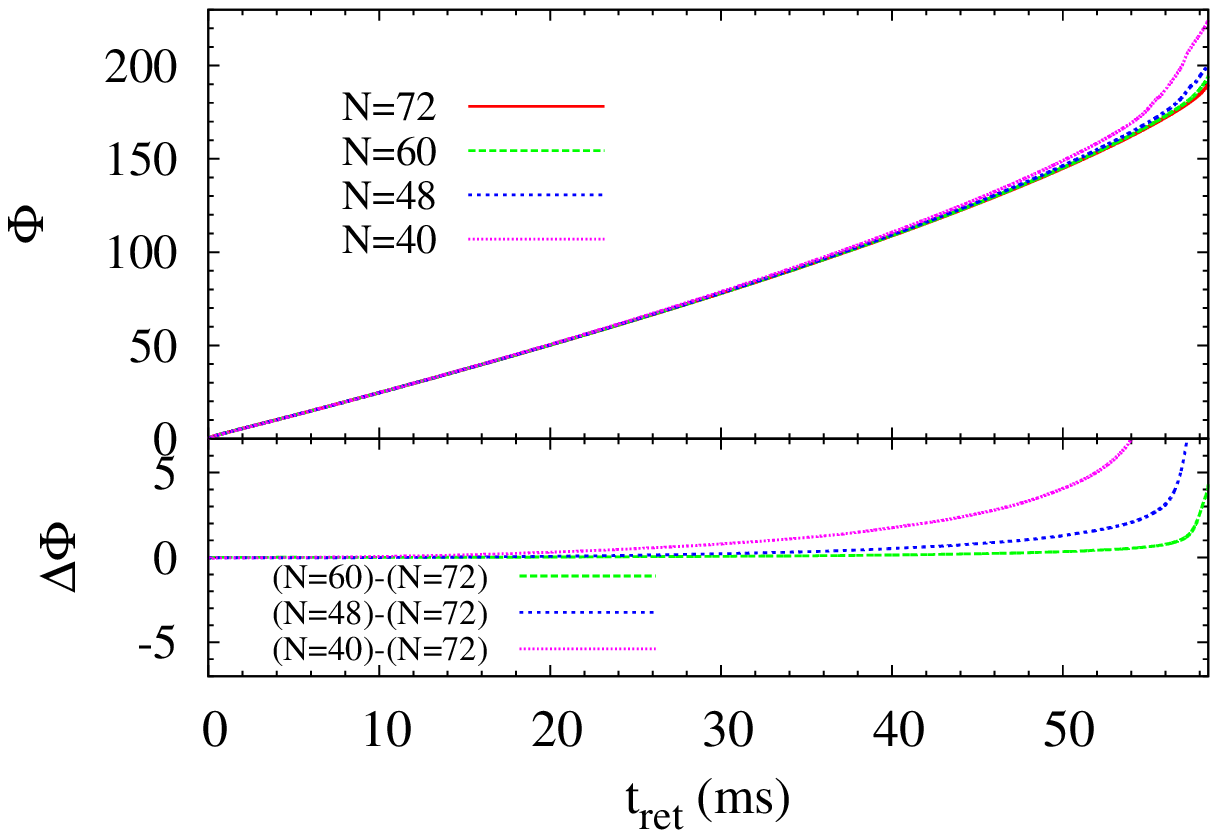}~~~
\includegraphics[width=84mm]{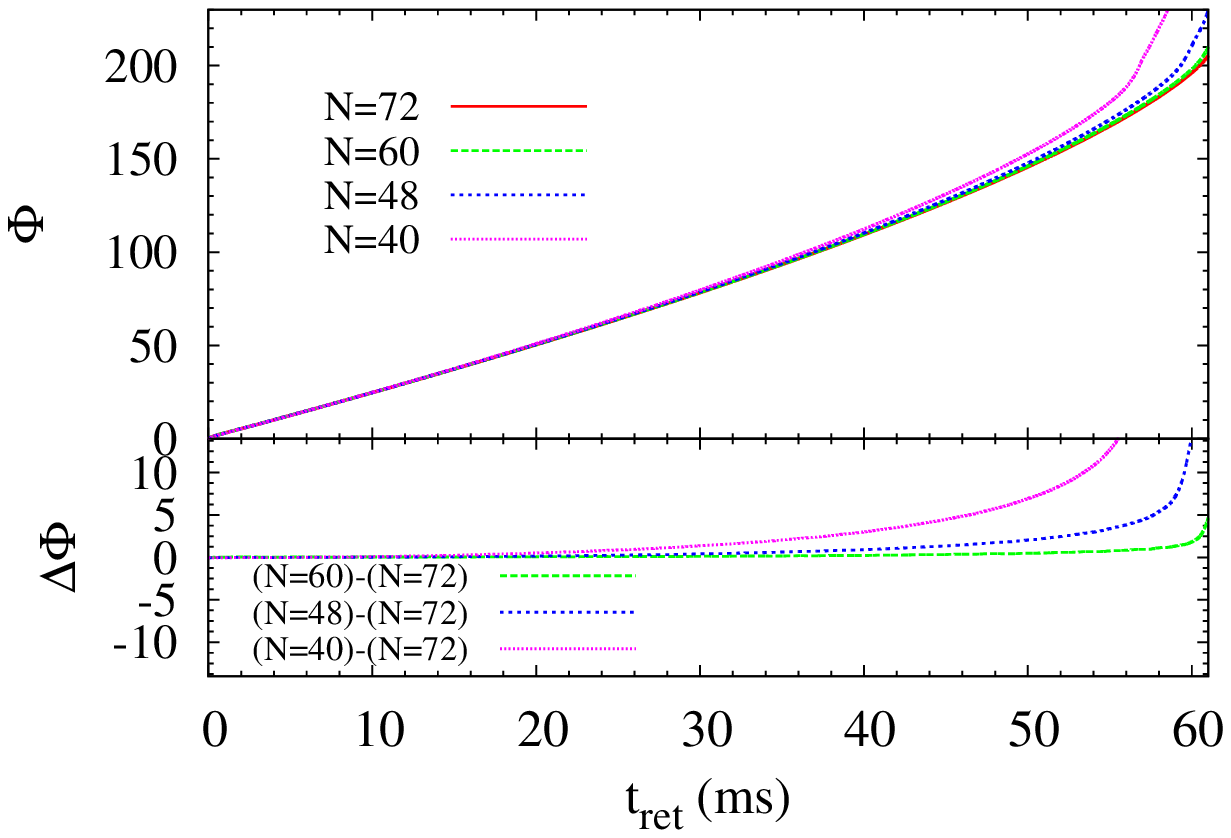}\\
\includegraphics[width=84mm]{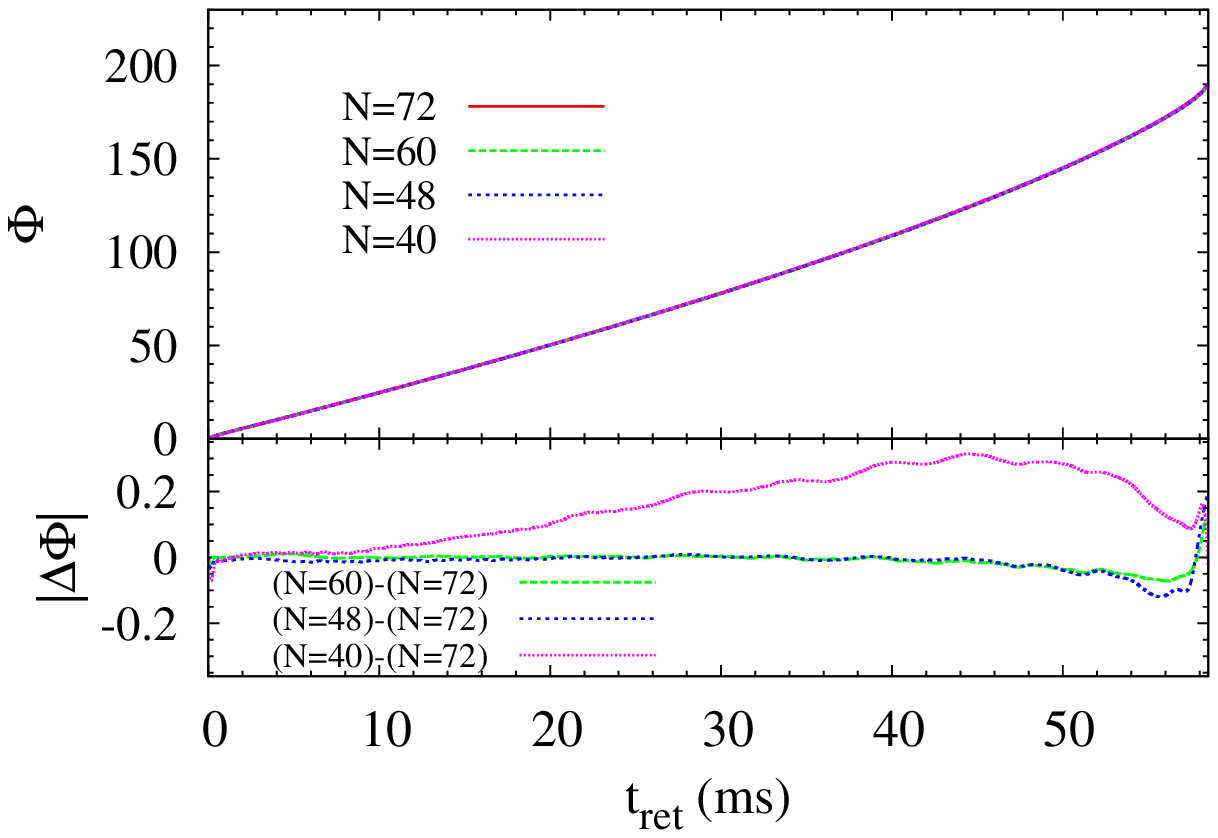}~~~
\includegraphics[width=84mm]{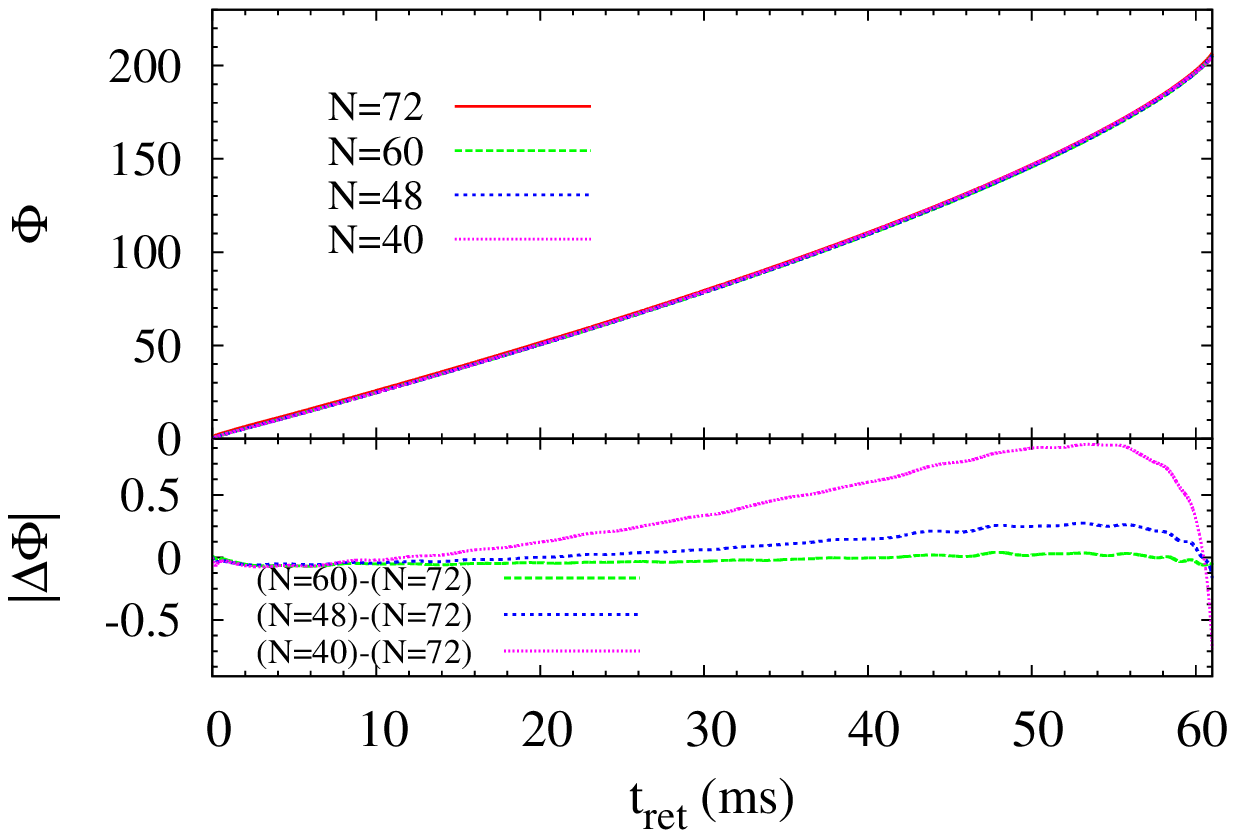}
%\vspace*{-4mm}
\caption{The gravitational waveforms and the evolution of the
  gravitational-wave phase for four different grid-resolution runs
  with the H4 EOS (left three panels) and with the APR4 EOS (right
  three panels). $N$ indicates the grid resolution, $\Delta x_8
  \propto N^{-1}$.  The upper panels show the gravitational waveforms
  for three different grid resolutions.  The middle panels show the
  pure numerical wave phases and the bottom panels show the results
  obtained after the stretching of time and phase according to the
  convergence property (for $N=40$, 48, and 60). The lower plots in
  middle and bottom panels show the phase disagreement between the
  purely numerical wave phase for $N=72$ and the lower-resolution
  results.  Note that for $N=72$, $t_{\rm mrg}=58.43$\,(ms) for the H4
  EOS and $t_{\rm mrg}=61.08$\,(ms) for the APR4 EOS.
\label{fig2}}
\end{center}
\end{figure*}

%%%%%%%%%%%%%%%%%%%%%%%%%%%%% VARIOUS N 

Next, we consider the resolution extrapolation for the limit $\Delta
x_8 \rightarrow 0$. For this task, numerical simulations have to be
performed for more than three grid resolutions. In this study, we
performed four simulations for each model employing four different
grid resolutions (cf.~Table~\ref{tab1} for the finest grid spacing,
$\Delta x_8$, for each run).  For each run, we extracted the numerical
waveform at $r_0=200m_0$ and then performed the extrapolation of $r_0
\rightarrow \infty$ as described in Eq.~(\ref{eq:nakano}).

%%%% FIGURE: MERGER TIME vs Delta x

We then need to perform an extrapolation procedure of taking the zero
grid-spacing limit for obtaining an approximately exact solution.  For
this procedure, we first analyze the relation of the time to the
merger, $t_{\rm mrg}$, as a function of $\Delta x_8$ following
Ref.~\cite{HKS2013}. Here, the merger time, $t_{\rm mrg}$, is defined
as the time at which the maximum value of $|\Psi_4^{2,2,\infty}(t_{\rm
  ret})|$ is recorded. Then, it is found that $t_{\rm mrg}$ converges
to an unknown exact value at $\sim 4$th order (see below for more
detailed analysis). $t_{\rm mrg}$ is larger for the better grid
resolutions because for the lower grid resolutions, the numerical
dissipation is larger and the inspiraling process is spuriously
accelerated. This numerical error is universally present for finite
values of $\Delta x_8$; namely, for any inspiraling stage in any
numerical simulations, the error is always present. For obtaining the
``exact'' waveform, thus, we always need an extrapolation
procedure. Then, the next question is how to extrapolate the waveform
for the limit $\Delta x_8 \rightarrow 0$. We propose the following
method in this study.

We first determine the gravitational waveform and time evolution of
the angular frequency as functions of $t_{\rm ret}$ by integrating
$\Psi_4^{l,m,\infty}(t_{\rm ret})$ for each raw numerical
data. Here, the gravitational waveform for each multipole mode
satisfies [see Eq.~(\ref{eq:hpsi4})]
\beq 
\ddot h^{l,m}:=\ddot h_+^{l,m}-i \ddot
h_\times^{l,m}=-\Psi_4^{l,m,\infty}(t_{\rm ret}).  
\eeq
$h^{l,m}$ is obtained by the double time integration of
$\Psi_4^{l,m,\infty}$.  For this procedure, we employ the method of
Ref.~\cite{RP2011}, written as
\beqn 
h^{l,m}(t_{\rm ret})=\int d\omega {\Psi_4^{l,m,\infty}(\omega)
  \over {\rm max}(\omega, \omega_{\rm cut})^2} \exp(i\omega t_{\rm
  ret}),~~
\label{eq:h}
\eeqn
where $\Psi_4^{l,m,\infty}(\omega)$ is the Fourier transform of
$\Psi_4^{l,m,\infty}(t_{\rm ret})$ and $\omega_{\rm cut}$ is chosen to
be $1.6\Omega_0$. (Note that at the initial stage, the value of
$\omega$ is $2 \Omega_0 > \omega_\mathrm{cut}$).  We recall again that
in this paper we pay attention only to $l=|m|=2$ modes because these
are the dominant modes in particular for the equal-mass
binaries. Then, from Eq.~(\ref{eq:h}), we determine the evolution of
the amplitude, i.e., $A^{l,m}=|h^{l,m}|$ as a function of $t_{\rm
  ret}$.

Using Eq.~(\ref{eq:h}), we can also define the evolution of the 
angular frequency as
\beqn
\omega(t_{\rm ret}):={|\dot h^{2,2}| \over |h^{2,2}|},
\eeqn
and then, the evolution of the gravitational-wave phase is calculated
by
\beqn
\Phi(t_{\rm ret}):=\int^{t_{\rm ret}} dt' \, \omega(t'). 
\eeqn
Now, using $A^{2,2}$ and $\Phi$, the quadrupole gravitational waveform 
can be written by
\beqn h^{2,2}(t_{\rm ret})=A^{2,2}(t_{\rm ret})\exp\left[i\Phi(t_{\rm
ret})\right].  \eeqn

Figure~\ref{fig2} plots the resulting gravitational waveforms and the
evolution of $\Phi$ obtained in the simulations with different grid
resolutions for the models with H4 (left) and APR4 EOS (right).  The
upper panels plot the gravitational waveforms and these show that the
merger time is earlier for the poorer grid resolutions.  The middle
panels plot the integrated wave phases for the pure numerical results
with no reprocessing. These show that the phase evolution is
spuriously faster for the poorer grid resolutions. However, we already
know that the merger time converges approximately at 4th-order. Taking
into account this fact, we stretch the time axis for the gravitational
waveform by an appropriate factor as $t \rightarrow \eta t$ where
$\eta (> 1)$ is the constant stretching factor. This factor should be
larger for the results of the poorer grid resolutions.  Here, this
reprocessing is performed in the same manner as in~\cite{HKS2013}:
$t_{\rm ret}$ and $\Phi$ are modified as $t_{\rm ret} \rightarrow \eta
t_{\rm ret}$ and $\Phi \rightarrow \eta \Phi$. We will show that the
phase evolution matches very well among the waveforms with different
grid resolutions after this scaling performed in terms of this single
parameter $\eta$. Later, $\eta$ will be also used for determining the
convergence order and for obtaining the resolution-extrapolated
waveform. 

As a first step for this stretching procedure, we have to determine
the values of $\eta$. As the first substep, we carry out a procedure
for finding the minimum value of the following integral
\beqn
I&=&\min_{\eta',\phi}\int_{t_i}^{t_f} dt_{\rm ret} \left|A_2^{2,2}(\eta' t_{\rm ret}) 
\exp\left[i \eta' \Phi_2(\eta' t_{\rm ret})+i \phi\right] \right. \nonumber \\
&& \hskip 2.1cm - \left. 
A_1^{2,2}(t_{\rm ret}) \exp\left[i \Phi_1(t_{\rm ret})\right] \right|^2,
\label{corre}
\eeqn
where $A_1^{2,2}$ and $\Phi_1$ are, respectively, the amplitude and
integrated phase of the gravitational waveform for the best-resolved
run ($N=72$) and $A_2^{2,2}$ and $\Phi_2$ are those for less-resolved
runs. The free parameters, $\eta'$ and $\phi$, are varied for a wide
range and from 0 to $2 \pi$, respectively, to search for the possible
minimum value of $I$. $t_i$ and $t_f$ are chosen to be 5\,ms and
$t_{\rm mrg}$ of the best resolved run, respectively. Here, the reason
for choosing $t_i=5$\,ms is that for their early stage with $t_{\rm
  ret} \alt 5$\,ms, the numerical waveforms have a relatively large
modulation in amplitude and phase due to junk radiation.

We find for our present simulation results that for the second-finest,
third-finest, and poorest resolution runs, $\eta'=1.00646$, 1.02241,
and 1.06000 for the H4 EOS and $\eta'=1.00650$, 1.02931, and 1.09118
for the APR4 EOS.  The mismatched factors, respectively, are
$I/I_0=7.4\times 10^{-6}$, $2.3 \times 10^{-5}$, and $1.4 \times
10^{-4}$ for the H4 EOS and $I/I_0=7.4\times 10^{-6}$, $1.1 \times
10^{-4}$, and $1.4 \times 10^{-3}$ for the APR4 EOS. Here, we define
\beq
I_0:= \int_{t_i}^{t_f} dt_{\rm ret} \left|A_1^{2,2}(t_{\rm ret})\right|^2.  
\eeq
The cross correlation of two waveforms is approximately estimated as
$1-\sqrt{I/2I_0}$. This implies that the cross correlation between the
waveforms of the best-resolved run and reprocessed less-resolved runs
are approximately 99.9\%, 99.8\%, 99.4\% for the H4 EOS and 99.9\%,
99.5\%, and 98.2\%, respectively. This shows that the accuracy is not
very good in the low-resolution runs for the APR4 EOS, for which the
compactness is larger than that for the H4 EOS, and hence, a finer
grid resolution would be necessary for a well-resolved simulation.
For both EOS, the reprocessed waveforms in the poorest-resolution run
are found to be not very accurate, and hence, in the following, we
will perform a convergence study employing the waveforms of the
first-, second-, third-resolved runs (labeled by $N=72$, 60, and 48,
respectively).

The bottom panels of Fig.~\ref{fig2} show the results obtained for
this time-stretching procedure.  It is found that four curves of
$\Phi$ originally with different grid resolutions approximately
overlap with each other. In particular, the degree of the overlapping
is quite good between the finest and second-finest runs (see the
difference of the integrated phase shown in the lower plot of the
bottom panels of Fig.~\ref{fig2}): For both EOS, the disagreement of
$\Phi$ for these reprocessed data is much smaller than 0.1 radian
except for the final moment of the last orbits, at which the
disagreement steeply increases: however it is at most $\sim
0.2$~radian. This suggests that the time stretching method can be used
for obtaining the extrapolated waveform for $\Delta x_8 \rightarrow 0$
if we accept the error of the integrated phase up to $\sim
0.2$~radian.

\begin{figure*}[t]
\begin{center}
\includegraphics[width=84mm]{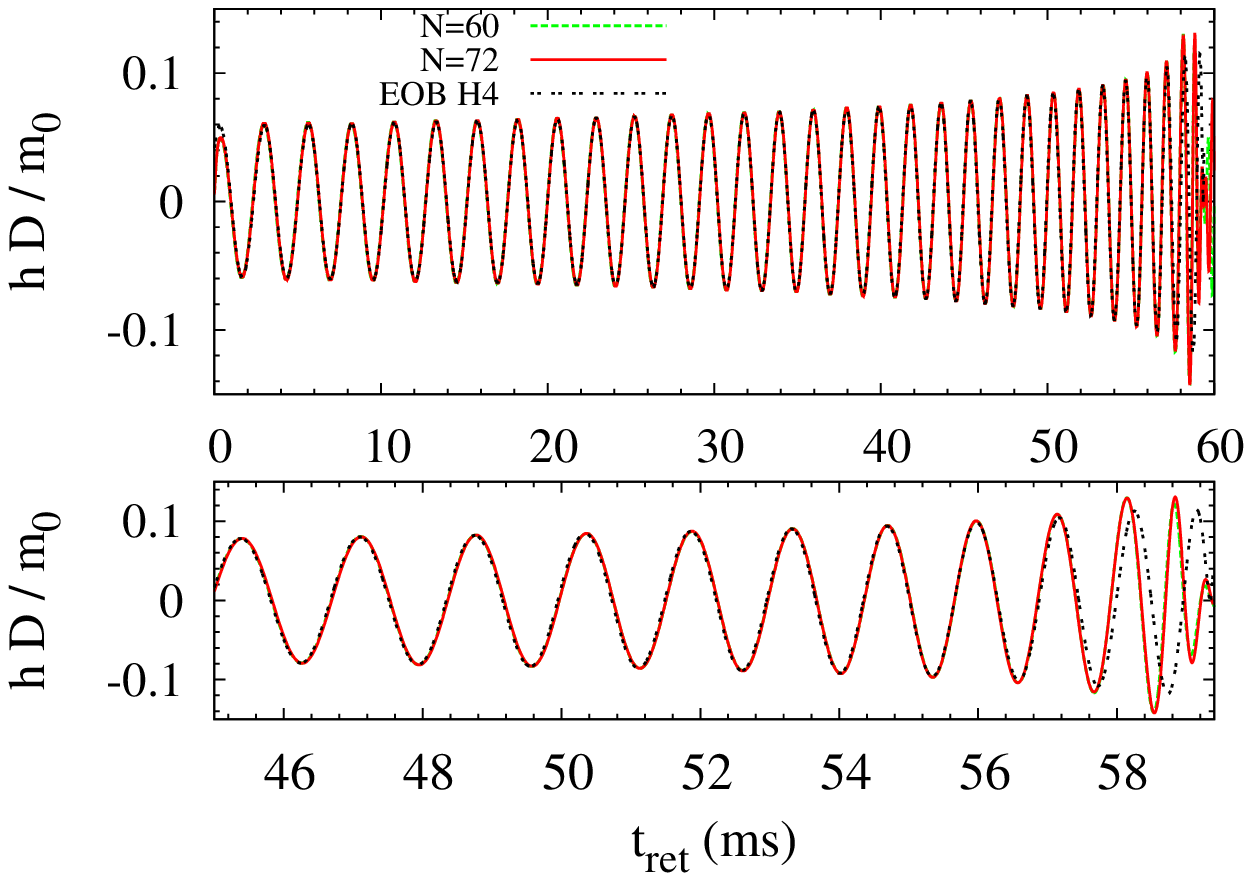}
\includegraphics[width=84mm]{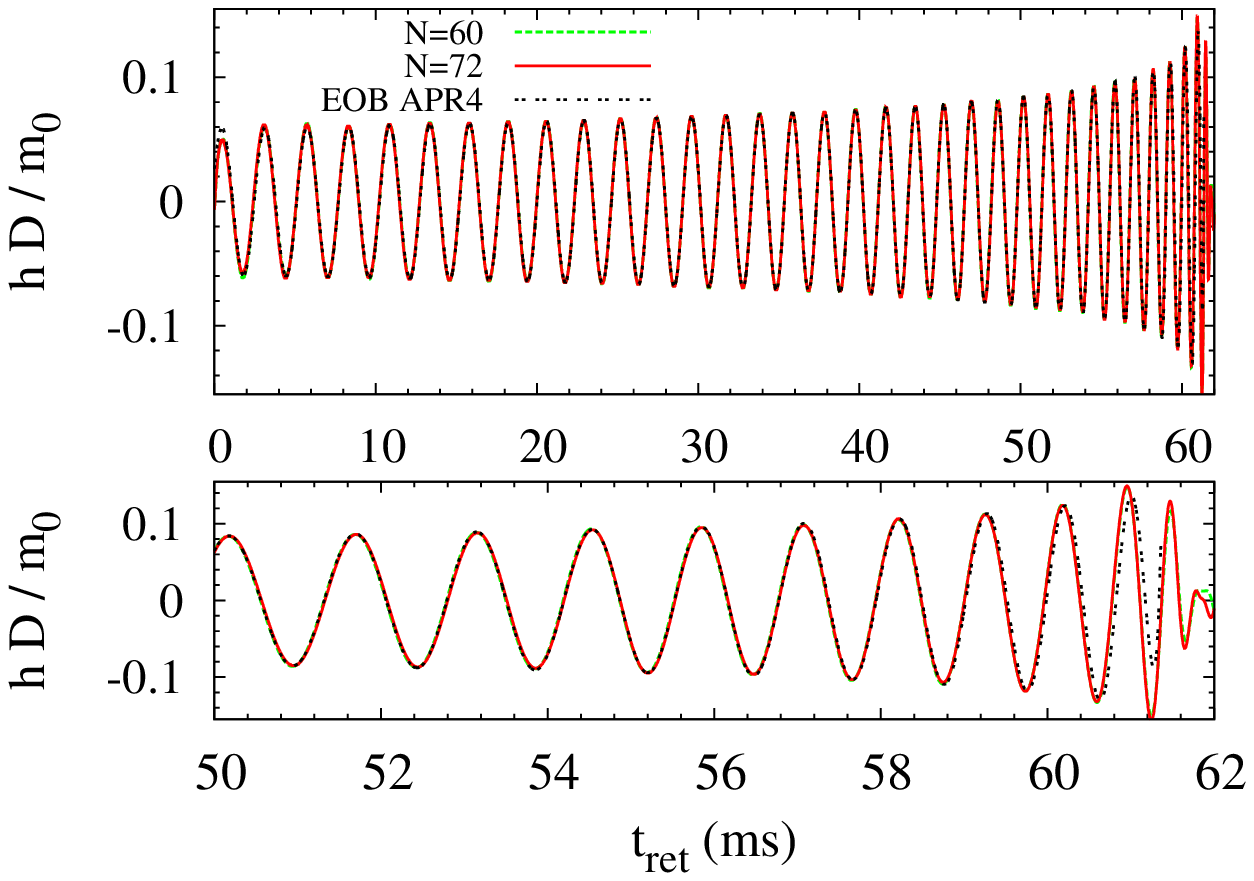}\\
\includegraphics[width=80mm]{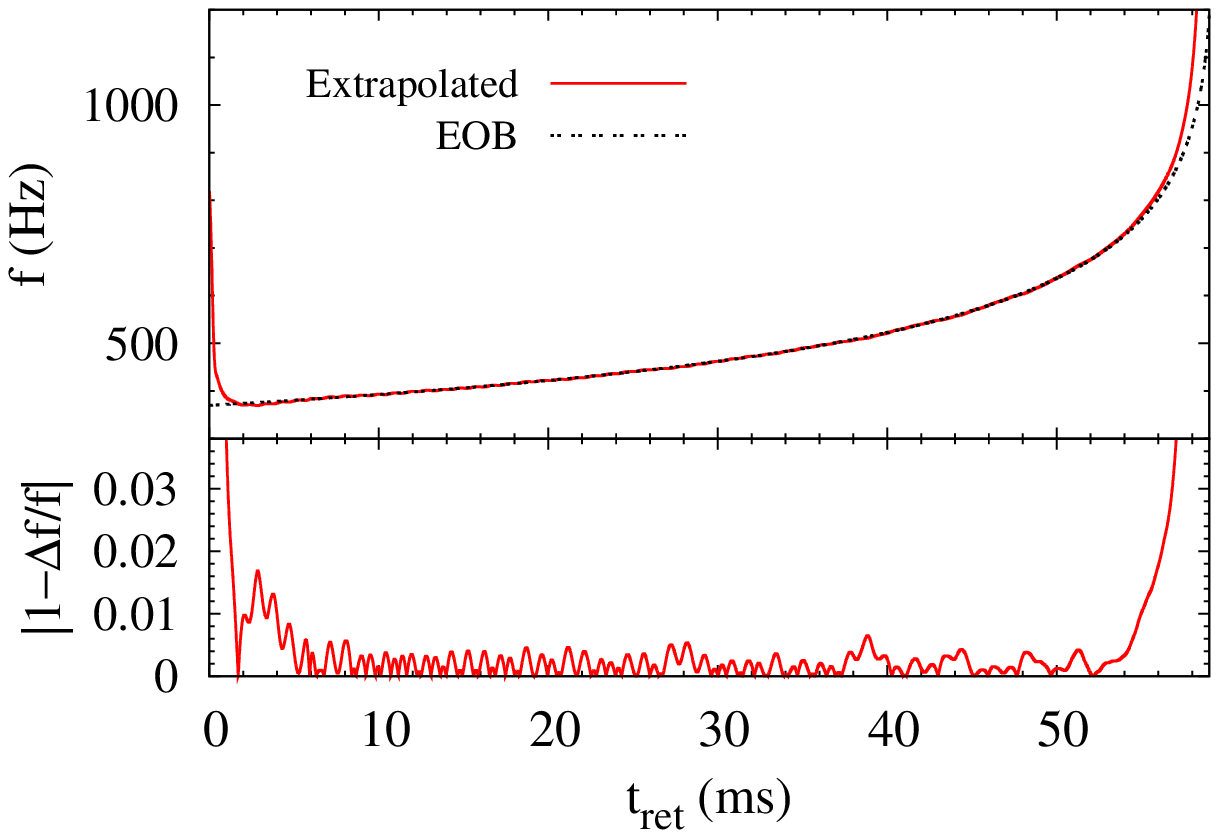}~~~~~
\includegraphics[width=80mm]{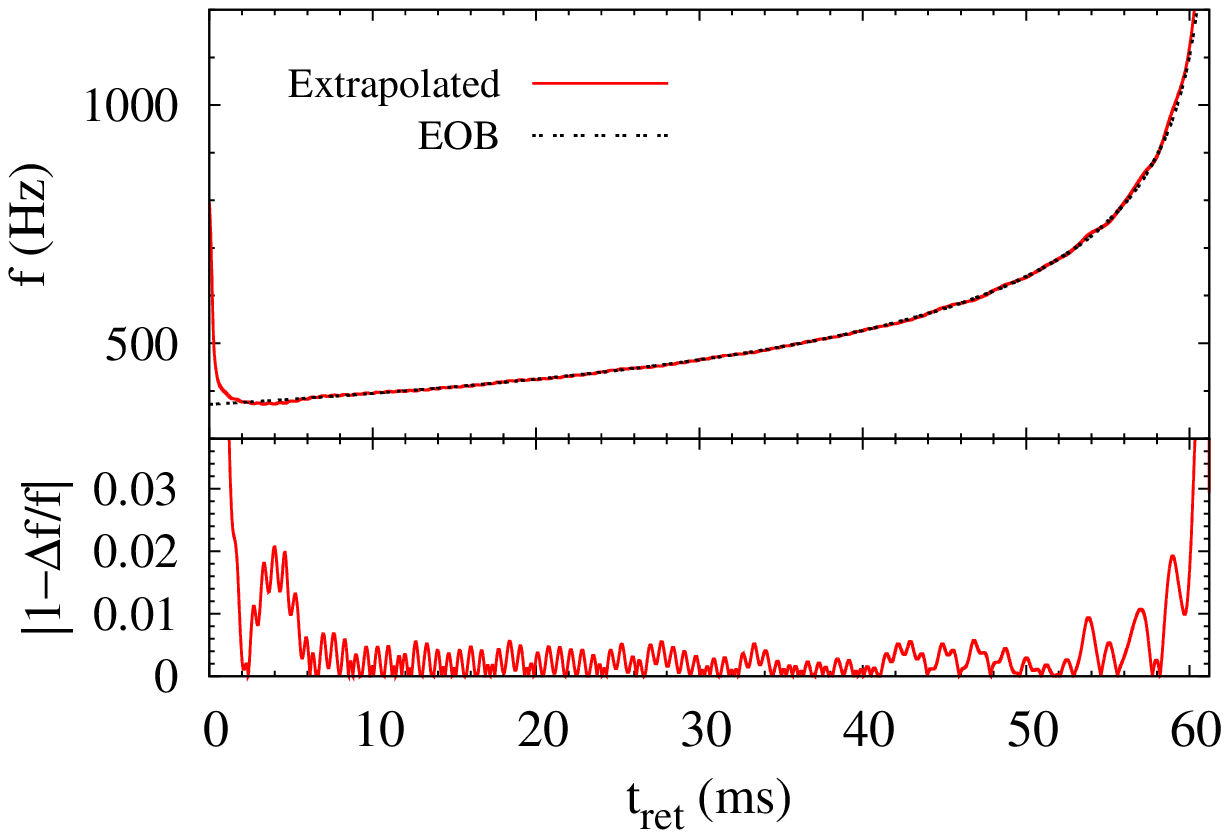}\\
\includegraphics[width=80mm]{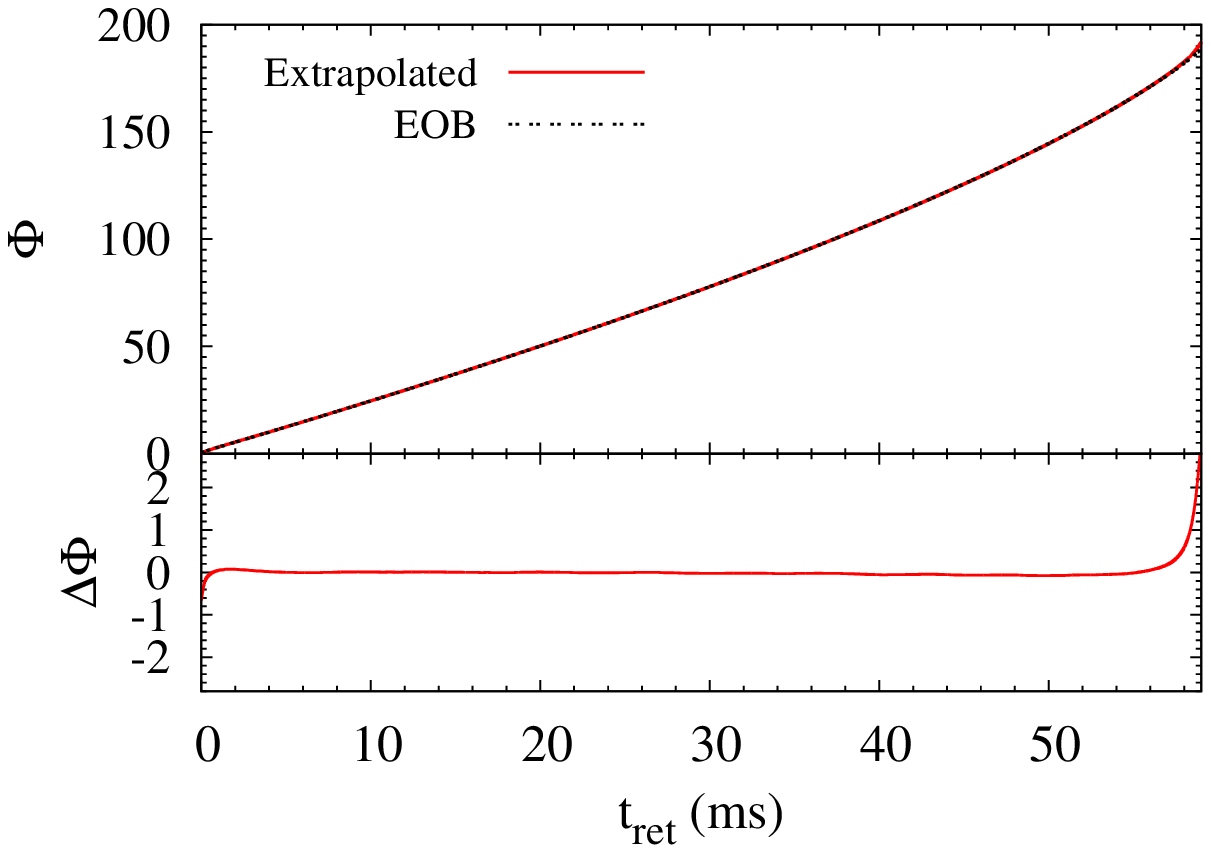}~~~~~
\includegraphics[width=80mm]{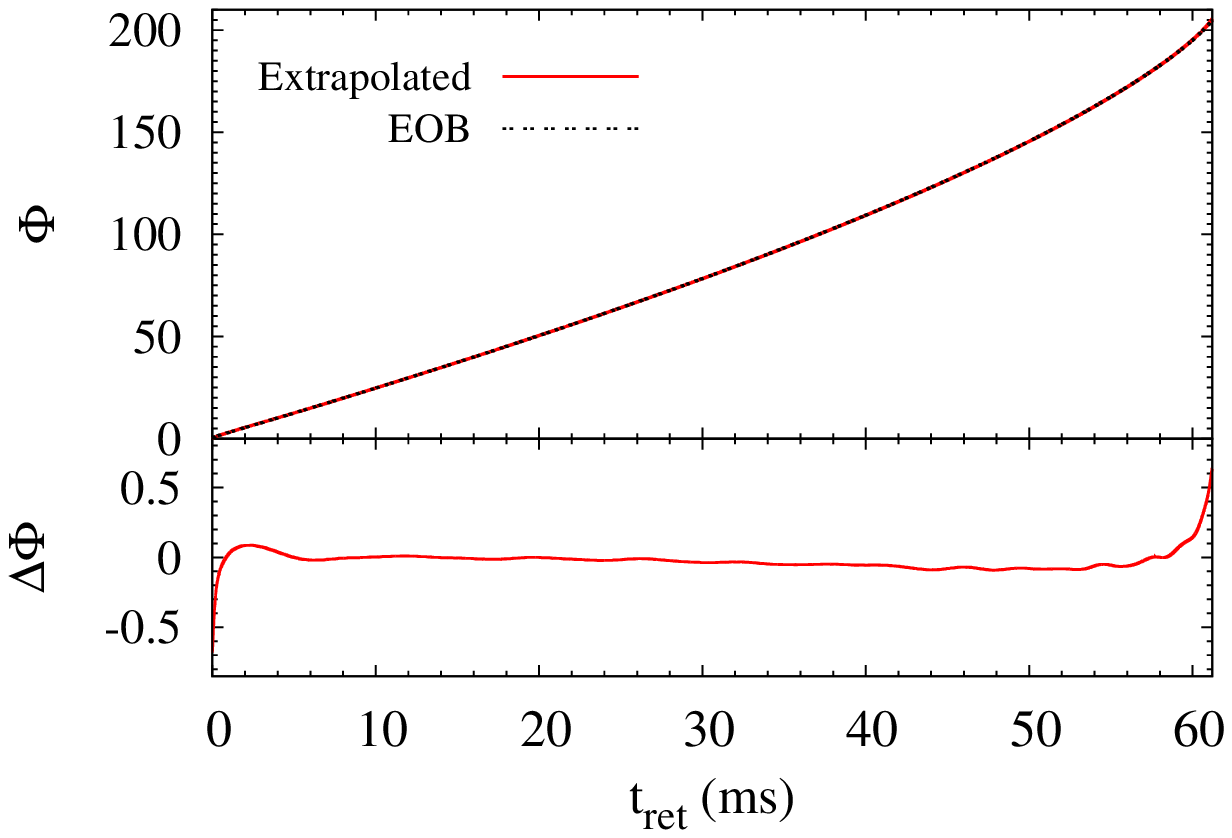}
%\vspace*{-4mm}
\caption{The extrapolated gravitational waveform and related
  quantities for the models with the H4 (left) and APR4 EOS
  (right). Top: The extrapolated waveforms for the best-resolved
  ($N=72$) and second-best-resolved ($N=60$) runs are plotted (two
  waveforms overlap quite well with each other and we cannot
  distinguish them in the figure).  The waveform by an
  effective-one-body calculation is plotted together. The lower panels
  focus on the late inspiral waveforms.  Middle: The extrapolated
  gravitational-wave frequency. In the lower panel of this, the
  absolute difference between the extrapolated result (with $N=72$)
  and EOB result is shown. Bottom: The extrapolated gravitational-wave
  phase. In the lower panel of this, the difference between the
  extrapolated result and EOB result is shown. We aligned the phases
  of the extrapolated and EOB waveforms at $t_{\rm ret}=5$\,ms.
\label{fig3}}
\end{center}
\end{figure*}

We next try to obtain an extrapolated waveform for $\Delta x_8
\rightarrow 0$ by using the time stretching method for the well
resolved models. For this procedure, we have to determine the order of
the convergence appropriately.  In the above, we found that the
numerical waveform in the poorest run is not very reliable even after
the reprocessing. Thus, we determine the order of the convergence from
the three better-resolved runs. (Note that if we employ the
poorest-resolved waveforms for determining it, the order of the
convergence is spuriously overestimated.)  Using the values of $\eta'-1$,
the order of the convergence, $p$, is determined from
\beqn
{(72/48)^p-1 \over (72/60)^p-1}=\left\{
\begin{array}{ll}
\displaystyle {0.02241 \over 0.00646} & {\rm for~H4},\\
& \\
\displaystyle {0.02931 \over 0.00650} & {\rm for~APR4},\\
\end{array}
\right.
\eeqn
which give $p \approx 3.42$ and $5.10$, for the H4 and APR4 EOS,
respectively. This indicates that the stretching factor for the
best-resolved run to reproduce the limiting waveform with $\Delta x_8
\rightarrow 0$ is $\eta \approx 1.00746$ and 1.00424 for the H4 and
APR4 EOS, respectively.  This implies that for these models, the exact
merger time would be $t_{\rm mrg} \approx 58.87$\,ms and 61.34\,ms,
respectively, whereas they were $58.43$\,ms and 61.08\,ms for the
best-resolved run. Namely, the error in the merger time is still much
larger than 0.1\,ms even for the best-resolved run: For obtaining the
waveforms of the error in the merger time smaller than 0.1\,ms, a
simulation with $N \agt 100$ would be necessary.  We note that if we
extrapolate the value of $\eta$ for the best-resolved runs assuming
the third- and fourth-order convergences of $\eta'$, the value of
$\eta$ becomes, respectively, $1.00887$ and $1.00601$ for the H4 EOS
and $1.00893$ and $1.00605$ for the APR4 EOS.  For the hypothetical
fourth-order convergence, the predicted merger time would be $t_{\rm
  mrg}=58.78$\,ms for the H4 EOS and $61.45$\,ms for the APR4
EOS. Thus, it is safe to keep in mind that the extrapolated merger
time still has an error of $\sim 0.1$\,ms due to the uncertainty in
$p$.  Since the merger time is $\sim 60$\,ms and total
gravitational-wave phase is $\sim 200$\,radian for both EOS, we should
keep in mind the phase error of $200 \times (0.1/60) \sim
0.3$\,radian.

\section{Comparison between numerical-relativity and effective-one-body
 waveforms}

Figure~\ref{fig3} plots the extrapolated gravitational waveforms, the
associated frequency, and the integrated gravitational-wave phase. For
comparison, we plot the results by an effective-one-body (EOB)
approach~\cite{DN2010,damour12,BDF2012} (see appendix A for the EOB
formalism that we employ in this work). To align the time and phase of
the numerical and EOB waveforms, we first calculate a correlation like
Eq.~(\ref{corre}) for $5\,{\rm ms} \leq t_{\rm ret} \leq 20\,{\rm ms}$
between the numerical and EOB waveforms, varying the time and phase of
the EOB waveform. These parameters are determined by searching for the
set of the values that give the minimum of this integral.

Figure~\ref{fig3} shows that up to $f \sim 700$\,Hz (at $t_{\rm ret}
\approx 54$\,ms), the EOB result well reproduces the extrapolated
waveforms for both H4 and APR4 EOS: In particular for the APR4 EOS for
which the compactness is large and the tidal deformability is small,
the agreement is quite good. For both EOS, the error in the frequency
is smaller than 1\% and the phase error is smaller than 0.1~radian for
$f \alt 700$\,Hz (with $t_{\rm ret} \geq 5$\,ms).  However, for the
last a few cycles, the agreement between the extrapolated and EOB
waveforms becomes poor. Here, note that this disagreement cannot be
explained by the error in the numerical waveform, because we have
already estimated that the phase error in the numerical waveform would
be smaller than $\sim 0.3$ radian. The magnitude of the error is
larger for the H4 EOS.  The possible reason for this disagreement is
that in the current version of the EOB formalism, the tidal effects
are not fully taken into account (e.g., non-linear tidal effects and
non-stationary effects are not included). Namely, if the degree of the
tidal deformation becomes high, the approximation could be poor. 

In the final inspiraling stage for the model with the H4 EOS, the
neutron stars are significantly deformed, and the attractive force
associated with the tidal deformation is enhanced: The relative
fraction of the approaching velocity induced by the tidal effect to
that by other general relativistic effects such as
gravitational-radiation reaction is larger for the binary of
larger-radius neutron stars.  The missing tidal effects could give a
significant damage in the current version of the EOB formalism. By
contrast, for the model with the APR4 EOS, the agreement between the
extrapolated and EOB waveforms is quite good even at the last
orbit. The total phase error is smaller than $\sim 0.7$~radian, which
is comparable to that in the error associated with the uncertainty of
the extrapolation. This implies that for the binary of small-radius
neutron stars, the current version of the EOB formalism would be
already robust if we accept the phase error of $\sim 1$~radian (see
also Ref.~\cite{bernuzzi14}).

\begin{figure}[t]
\begin{center}
\includegraphics[width=84mm]{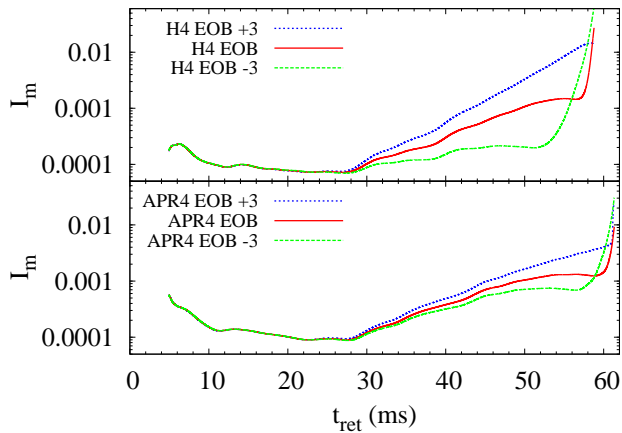}
%%\includegraphics[width=84mm]{fig4a.eps}\\
%%\includegraphics[width=84mm]{fig4b.eps}\\
%%\includegraphics[width=84mm]{fig4c.eps}
%%%\vspace*{-4mm}
%%\caption{The same as Fig.~\ref{fig3} but for comparison with the
%%  numerical and EOB waveform in which the EOB results are obtained
%%  using a slightly modified EOS.
\caption{The evolution of the mismatch, $I_m(t_{\rm ret})$, between the
  extrapolated waveform and EOB waveforms.  ``EOB\,$\pm 3$'' denotes
  that the EOB waveforms are employed artificially increasing or
  decreasing the neutron-star compactness by 3\%.
\label{fig4}}
\end{center}
\end{figure}

The missing tidal effects in the EOB formalism cannot be compensated
even if we artificially modify the value of the tidal deformability
(or compactness). Figure~\ref{fig4} plots the evolution for the degree
of the mismatch between the extrapolated waveform and the EOB
waveform. For comparison, we calculated the mismatch employing the EOB
waveforms in which the compactness of neutron stars is varied by $\pm
3\%$.  Here, the mismatch is defined by
\beqn
I_m(t_{\rm ret}):={1 \over 2}{(h-h_{\rm eob}|h-h_{\rm eob}) \over (h_{\rm eob}|h_{\rm eob})}, 
\eeqn
where 
\beq
(h_1|h_2):=\int_{t_i}^{t_{\rm ret}} h_1(t_{\rm ret}') h_2^*(t_{\rm ret}') dt_{\rm ret}'. 
\eeq
Again, $t_i$ is chosen to be 5\,ms. Here, $h$ and $h_{\rm eob}$ denote an  
extrapolated waveform and a waveform by the EOB formalism, 
respectively. We note that the following relation is approximately 
satisfied for small values of $I_m(t_{\rm ret})$: 
\beqn
1-I_m(t_{\rm ret}) \approx {(h|h_{\rm eob}) \over \sqrt{(h|h)(h_{\rm eob}|h_{\rm eob})}}. 
\eeqn
From Fig.~\ref{fig4}, we first reconfirm that the degree of the
mismatch is steeply increased for the last inspiral orbit.  This
indicates that the tidal effect would not be sufficiently taken into
account in the current version of the EOB formalism, although for
other inspiral orbits, the performance of the EOB formalism appears to
be quite good. It is also found that the extrapolated waveforms cannot
be accurately reproduced even if we simply change the tidal
deformability: If its value is artificially increased, the phase
evolution is accelerated, and as a result, the mismatch is increased
in an earlier inspiral stage. If it is artificially decreased, the
merger is delayed, and as a result, the mismatch is badly increased
near the last orbit.  This suggests that a tidal effect, which is not
included, should be taken into account for improving the performance
of the EOB formalism.

\section{Summary}

We presented our latest numerical results of longterm simulations for
the inspiraling binary neutron stars of equal mass.  By a careful
resolution study and extrapolation procedure, we obtain an accurate
waveform: The estimated total phase error is smaller than $\sim 0.3$
radian for the total integrated phase of $\sim 200$ radian and the
maximum error in the wave amplitude is smaller than 3\%. Using these
accurate waveforms, we calibrated the waveforms derived by the latest
EOB formalism.  We show that for a binary of compact neutron stars
(with their radius 11.1\,km), the waveform by the EOB formalism agrees
quite well with the numerical waveform so that the total phase error
is smaller than 1 radian. By contrast, for a binary of less compact
neutron stars (with their radius 13.6\,km), the EOB and numerical
waveforms disagree with each other in the last a few wave cycles,
resulting in the total phase error of $\sim 3$ radian.  We infer that
this is due to the missing of some tidal effect such as nonlinear
tidal effect in the current version of the EOB formalism, which should
be taken into account for improving its performance.

In this work, we employed only two representative EOS and a binary of
particular mass. For systematically improving the EOB formalism, we
have to derive waveforms of wider sets of EOS and binary mass.  We
plan to perform more simulations in the future work and to present a
larger number of the waveforms using the prescription developed in
this paper. 

\begin{acknowledgments}

This work was supported by Grant-in-Aid for Scientific Research
(24244028) of Japanese MEXT. KK is supported by JSPS Postdoctoral
Fellowship for Research Abroad.

\end{acknowledgments}

\appendix
\section{Effective one body formalism}

In this work, we employ an EOB formalism for inspiraling binary
neutron stars, which is described in Ref.~\cite{bernuzzi14}. The base
point-particle dynamics for this EOB formalism is calibrated by the
latest binary-black-hole merger simulations~\cite{DNB2013} and the
tidal effects are taken into account based on the prescription of
Refs.~\cite{BDF2012,damour12,BD2014}.  Here, we briefly review this 
type of the EOB formalisms and describe our choice. 

We consider a binary system composed of stars $A$ and $B$ with mass of
$M_{A}$ and $M_{B}$.  The EOB effective metric is defined by
\begin{eqnarray}
ds_{\rm{eff}}^{2} = -A(r)dt^{2}+\frac{D(r)}{A(r)}dr^{2}+r^{2}\left(d\theta^{2}
+\sin^{2}\theta d\phi^{2}\right), \nonumber \\
\end{eqnarray}
where $(r,\phi)$ are dimensionless coordinates and their canonical
momenta are $(p_{r},p_{\phi})$.  We replace the radial canonical
momentum $p_{r}$ with the canonical momentum $p_{r_{*}}$, where a
tortoise-like radial coordinate $r_{*}$ is given by
\begin{eqnarray}
\frac{dr_{*}}{dr}=\frac{\sqrt{D(r)}}{A(r)}.
\end{eqnarray}
Then, the binary dynamics can be described by the EOB Hamiltonian
\begin{eqnarray}
H_{\rm{real}}(r,p_{r_{*}},p_{\phi}) = Mc^{2}\sqrt{1+2\nu
  \left(\hat{H}_{\rm{eff}}-1\right)}\,,~~
\end{eqnarray}
where $\nu:=M_{A}M_{B}/M^{2}$, $M:=M_{A}+M_{B}$, and the effective
Hamiltonian is defined by
\begin{eqnarray}
\hat{H}_{\rm{eff}} =
\sqrt{p_{r_{*}}^{2}+A(r)\left(1+\frac{p_{\phi}^{2}}{r^{2}}+2\left(4-3\nu
  \right)\nu \frac{p_{r_{*}}^{4}}{r^{2}}\right)}\,\,.\nonumber \\
\end{eqnarray}

The potential $A(r)$ is decomposed into two parts as
\begin{eqnarray}
A(r) = A_{\rm{pp}}(r)+A_{\rm{tidal}}(r),
\end{eqnarray}
where $A_{\rm{pp}}(r)$ is the point-particle potential and
$A_{\rm{tidal}}(r)$ is the term associated with tidal effects.
The point-particle potential including up to the fifth PN terms is
\begin{eqnarray}
A_{\rm{pp}}(r)&=&
P_{5}^{1}\left[1-2u+2\nu u^{3}+\nu a_{4}u^{4}\right. \nonumber \\
&&~~~ \left. + \nu(a_{5}^{\rm c}(\nu)+a_{5}^{\rm ln}\ln u) u^{5} \right.\nonumber \\ 
&&~~~ \left. + (a_{6}^{\rm c}(\nu)+\nu a_{6}^{\rm ln} (\nu)\ln u)u^{6} \right], 
\end{eqnarray}
where $u := 1/r$, and $P^{1}_{5}$ denotes a $(1,5)$ Pad{\'e}
approximant.  Here, the following coefficients are analytically
known~\cite{barausse12,BD2013} 
\begin{eqnarray}
a_{4}& = &\frac{94}{3}-\frac{41}{32}\pi^{2},\\
a_{5}^{\rm c}(\nu) & = &
-\frac{4237}{60}+\frac{2275}{512}\pi^{2}+
\frac{256}{5}\ln 2+\frac{128}{5}\gamma \nonumber \\
& &  +\left(-\frac{211}{6}+\frac{41}{32}\pi^{2}\right)\nu ,\\
a_{5}^{\rm ln} & =& \frac{64}{5},\\
a_{6}^{\rm ln}(\nu) & =& -\frac{7004}{105} -\frac{144}{5}\nu,
%\begin{tabular}{l l l}
%$a_{4}$& $=$ &$\displaystyle\frac{94}{3}-\frac{41}{32}\pi^{2},$\\
%$a_{5}^{\rm c}(\nu)$ & $=$ &$
%\displaystyle-\frac{4237}{60}+\frac{2275}{512}\pi^{2}+
%\frac{256}{5}\ln 2+\frac{128}{5}\gamma$ \\
%& &  $\displaystyle+\left(-\frac{211}{6}+\frac{41}{32}\pi^{2}\right)\nu ,$\\
%$a_{5}^{\rm ln}$ & $=$& $\displaystyle\frac{64}{5},$\\
%$a_{6}^{\rm ln}(\nu)$ & $=$& $\displaystyle-\frac{7004}{105} -\frac{144}{5}\nu,$
%\end{tabular}
\end{eqnarray}
where $\gamma = 0.5772156 \dots$ is the Euler constant. Following 
Ref.~\cite{bernuzzi14}, we take the effective form of $a_{6}^{\rm
c}(\nu)$, with which results of binary-black-hole-merger simulations are
reproduced accurately, as
\begin{eqnarray}
a_{6}^{\rm c}(\nu) = 3097.3\nu^{2} -1330.6\nu + 81.38.
\end{eqnarray}

The contribution of tidal effects to the potential is written as
\begin{eqnarray}
A_{\rm{tidal}}(r) =- \sum_{l \geq 2} \left(
\kappa_{l}^{A}u^{2l+2}\hat{A}_{A}^{(l)}(u) + \left(A \leftrightarrow B
\right)\right),~~~
\end{eqnarray}
where $\hat{A}_{A}^{(l)}$ includes the PN tidal effects and
$\kappa_{l}^{A}$ is the tidal coefficients. Here, the subscripts $A$
and $B$ denote the stars $A$ and $B$.  In this work, we include the
tidal effects up to $l=4$.  The coefficient $\kappa_{l}^A$ is related to
the electric tidal Love number $k_{l}$ and the compactness $C$ as~(see
Table~\ref{tab1} for these values of the neutron stars studied in this
work)
\begin{eqnarray}
\kappa_{l}^{A} = 2\frac{M_{B}M_{A}^{2l}}{M^{2l+1}}\frac{k^{A}_{l}}{C^{2l+1}_{A}}.
\end{eqnarray}
The tidal potential up to the next-to-next-to-leading corrections is
\begin{eqnarray}
\hat{A}^{(l)}_{A}(u) = 1+\alpha_{A,1}^{(l)}u+\alpha_{A,2}^{(l)}u^{2}.\label{NNLO}
\end{eqnarray}
The coefficients are analytically known as \cite{BDF2012}
\begin{eqnarray}
\alpha_{A,1}^{(2)} &=& \frac{5}{2}X_{A},\\
\alpha_{A,2}^{(2)} &=& \frac{337}{28}X^{2}_{A}+\frac{1}{8}X_{A}+3,\\
\alpha_{A,1}^{(3)} &=& \frac{15}{2}X_{A}-2,\\
\alpha_{A,2}^{(3)} &=& \frac{110}{3}X_{A}^{2}-\frac{311}{24}X_{A}+\frac{8}{3},
%\begin{tabular}{l l l}
%$\alpha_{A,1}^{(2)}$ &$=$& $\displaystyle\frac{5}{2}X_{A}$,\\
%$\alpha_{A,2}^{(2)}$ &$=$& $\displaystyle\frac{337}{28}X^{2}_{A}+\frac{1}{8}X_{A}+3$,\\
%$\alpha_{A,1}^{(3)}$ &$=$& $\displaystyle\frac{15}{2}X_{A}-2$,\\
%$\alpha_{A,2}^{(3)}$ &$=$& $\displaystyle\frac{110}{3}X_{A}^{2}-\frac{311}{24}X_{A}+\frac{8}{3}$,\\
%\end{tabular}
\end{eqnarray}
where $X_{A}:=M_{A}/M$. 

Recently, the tidal EOB was improved using resummation
techniques~\cite{BD2014}.  We use the gravitational-self-force
informed $l=2$ tidal potential as
\begin{eqnarray}
\hat{A}^{(2)}_{A}(u) &=& 1+\frac{3u^{2}}{1-r_{\rm LR}u}
 + X_{A}\frac{\tilde{A}_{A}^{\rm (2)1SF}(u)}{(1-r_{\rm LR})^{7/2}}\nonumber \\
&&+X_{A}^{2}\frac{\tilde{A}_{A}^{\rm (2)2SF}(u)}{(1-r_{\rm LR})^{p}},
\end{eqnarray}
where $p$ is an unknown parameter in the range of $4\leq p < 6$ and
we set $p$ to be 4. 
$r_{\rm LR}$ is the light-ring orbit.
The forms of $\tilde{A}_{A}^{\rm (2)1SF}$ and $\tilde{A}_{A}^{\rm (2)2SF}$ are
\begin{eqnarray}
\tilde{A}_{A}^{\rm (2)1SF}(u) & = & \frac{5}{2}u(1-a_{1}u)(1-a_{2}u)\frac{1+n_{1}u}{1+d_{2}u^{2}},\\
\tilde{A}_{A}^{\rm (2)2SF}(u) & = & \frac{337}{28}u^{2},
\end{eqnarray}
where the numerical coefficients $(a_{1}, a_{2}, n_{1}, d_{2})$ are
found in Ref.~\cite{BD2014}. As in Ref.~\cite{bernuzzi14}, we use the
tidally corrected light ring orbit instead of $r_{\rm LR}=3$.  For
determining the value of $r_{\rm LR}$, we solve the following equation
numerically
\begin{eqnarray}
A(u_{\rm LR}) + \left. \frac{1}{2}u_{\rm LR}\frac{dA}{du}\right|_{u_{\rm LR}} = 0,
\end{eqnarray}
where the tidal part of the potential is included as Eq.~(\ref{NNLO})
and the value of $r_{\rm LR}$ for the binary neutron star models
employed in this work is shown in Table~I.  Finally, the potential
$D(u;\nu)$ is given by
\begin{eqnarray}
D(u;\nu)=\frac{1}{1+6\nu u^{2} + 2(23-3\nu)\nu u^{3}}.
\end{eqnarray}

For calculating the dynamics of the binary orbit under the potentials
described above, we solve the EOB Hamilton's equations
\begin{eqnarray}
 \frac{dr}{dt} &=& \frac{A(r)}{\sqrt{D(r)}}\frac{\partial H_{\rm{real}}}{\partial p_{r_{*}}},\\
 \frac{d\phi}{dt} &=& \frac{\partial H_{\rm{real}}}{\partial p_{\phi}},\\
\frac{dp_{r_{*}}}{dt} &=& -\frac{A(r)}{\sqrt{D(r)}}\frac{\partial H_{\rm{real}}}{\partial r},\label{hampr} \\
\frac{dp_{\phi}}{dt} &=& \mathcal{F}_{\phi}.
\end{eqnarray}
Note that we do not include the radial part of the radiation-reaction
force in Eq.~(\ref{hampr})~\cite{bernuzzi14} because we find this
choice advantageous for fitting the extrapolated gravitational
waveforms.  $\mathcal{F}_{\phi}$ is the radiation-reaction force given
by
\begin{eqnarray}
\mathcal{F}_{\phi} = -\frac{1}{8\pi \nu \omega}
\sum_{l=2}^{8}\sum_{m=1}^{l}\left(m\omega \right)^{2}|Rh_{lm}|^{2},
\end{eqnarray}
where $\omega = d\phi/dt$ and $h_{lm}$ denotes the multipolar waveforms.
Here, $h_{lm}$ is written as
\begin{eqnarray}
h_{lm}=h_{lm}^{0}+h_{lm}^{\rm{tidal,A}}+h_{lm}^{\rm{tidal,B}},
\end{eqnarray}
where $h_{lm}^{0}$ includes the inspiral and plunge waveforms given in
Ref.~\cite{DNB2013}, and $h_{lm}^{\rm{tidal,A}}$ and
$h_{lm}^{\rm{tidal,B}}$ are the tidal contributions due to the stars A
and B.  They are given by Eqs.~(A14)--(A17) of
Ref.~\cite{damour12}.

%%%%%%%%%%%%%%%%%%%%%%%%%%%

%%%%%%%%%%%%%%%%%%%%%

\end{document}